\documentclass[reqno,11pt]{article}
\usepackage{amsmath, amsthm, amscd,amsfonts, amssymb,hyperref,floatrow}
\usepackage{graphicx, color,dsfont,xcolor}

\usepackage{authblk}

\numberwithin{equation}{section}

\DeclareMathSymbol{\leqslant}{\mathalpha}{AMSa}{"36} 
\DeclareMathSymbol{\geqslant}{\mathalpha}{AMSa}{"3E} 
\DeclareMathSymbol{\eset}{\mathalpha}{AMSb}{"3F}     
\renewcommand{\leq}{\;\leqslant\;}                   
\renewcommand{\geq}{\;\geqslant\;}                   

\def \C{ \mathbb  C }
\newcommand{\R}{\mathbb{R}}

\newcommand{\N}{\mathbb{N}}

\def \P{ \mathbb P  }

\newcommand \be  {\begin{equation*}}
\newcommand \bea {\begin{eqnarray} \nonumber }
\newcommand \ee  {\end{equation*}}

\newcommand \ba  {\begin{align}}
\newcommand \ea  {\end{align}}

\definecolor{remi}{rgb}{0,0,0}

\begin{document}
\title{Invariant $\beta$-Wishart ensembles, crossover densities and asymptotic corrections to the Mar\v cenko-Pastur law}

\author[1,2]{Romain Allez}
\author[2]{Jean-Philippe Bouchaud} 
\author[3]{ Satya N. Majumdar}
\author[3]{Pierpaolo Vivo}

\affil[1]{Universit{\'e} Paris-Dauphine, Laboratoire CEREMADE, 
Place du Marechal de Lattre de Tassigny, 
75775 Paris Cedex 16 - France. }
\affil[2]{Capital~Fund~Management, 6--8 boulevard Haussmann, 75\,009 Paris, France.}
\affil[3]{Laboratoire de Physique Th\'{e}orique et Mod\`{e}les
Statistiques (UMR 8626 du CNRS), Universit\'{e} Paris-Sud,
B\^{a}timent 100, 91405 Orsay Cedex, France. }

\maketitle

\begin{abstract}
We construct a diffusive matrix model for the $\beta$-Wishart (or Laguerre) ensemble for 
general continuous $\beta\in [0,2]$, which preserves invariance under the 
orthogonal/unitary group transformation. 
Scaling the Dyson index $\beta$ with the largest size $M$ of the data matrix as 
$\beta=2c/M$ (with $c$ a fixed positive constant), we obtain a family of spectral 
densities parametrized by $c$. As $c$ is varied, this density interpolates continuously 
between the Mar\v cenko-Pastur ($c\to \infty$ 
limit) and the Gamma law ($c\to 0$ limit).  
Analyzing the full Stieltjes transform (resolvent) equation, we obtain as a 
byproduct the correction to the Mar\v cenko-Pastur density in the bulk up to order $1/M$ 
for all 
$\beta$ and up to order $1/M^2$ for the particular cases $\beta=1,2$.
\end{abstract}

\section{Introduction}

The theory of matrices with random entries, originally devised as a tool to understand and 
predict the spectra of heavy nuclei for which a detailed account of the interactions 
between particles is too complicated, has seen a spectacular resurgence of interest in 
recent years, with a number of unexpected and often surprising applications (see 
\cite{Handbook,AGZ,Bai,Forrester,Verdu} for a recent overview). While Wigner and Dyson are 
usually 
regarded as the pioneers in the field, John Wishart had already introduced random 
matrices in 1928 in his studies of multivariate populations \cite{wis}. The Wigner-Dyson 
(Gaussian) and Wishart ensembles (together with a few others) lie at the core of the 
\emph{classical} world of invariant matrices, characterized by the following main 
features: 

\begin{enumerate} 

\item The joint probability distribution (jpd) of matrix entries, collectively 
denoted by $P[\mathbf{X}]$, remains unaltered if one performs a similarity transformation 
$\mathbf{X}\to \mathbf{U X U}^{-1}$, with $\mathbf{U}$ being orthogonal (real symmetric 
$\mathbf{X}$), unitary (complex hermitian $\mathbf{X}$) or symplectic (quaternion 
self-dual $\mathbf{X}$) matrix. As a consequence, the eigenvectors of such matrices are 
Haar (uniform) distributed in their respective groups.

\item The joint distribution of the $N$ real eigenvalues $P(\lambda_1,\ldots,\lambda_N)$ 
can be generically written in the Gibbs-Boltzmann form,

\begin{equation}
P(\lambda_1,\ldots,\lambda_N)=
\frac{1}{Z_N} \exp\left(- \mathcal{H}(\lambda_1,\dots,\lambda_N)\right)\label{jpdeig}
\end{equation}

with the Hamiltonian $\mathcal{H}(\lambda_1,\dots,\lambda_N)$ given by:

\begin{equation}
\mathcal{H}(\lambda_1,\dots,\lambda_N) = \sum_{i=1}^N V(\lambda_i)- 
\beta \sum_{j< k}\ln |\lambda_j - \lambda_k| \label{ham}
\end{equation}
where $V(x)$ a confining potential that depends on the precise form of the joint 
distribution of matrix entries 
$P[\mathbf{X}]$. For example, if the entries of $\mathbf{X}$ are independent, the only 
allowed potential is quadratic $V(x)=\beta x^2/2$, which correspond to the Gaussian 
ensembles. If correlations among the entries are allowed, then different potentials (all 
corresponding to rotationally invariant weights) are possible. For example, 
in the Wishart case, $V(x)=\infty$ for $x<0$ (so that all the eigenvalues
are non-negative) and
$V(x)= x/(2\sigma^2) -\alpha \log x$ for $x\ge 0$.

The normalization constant $Z_N$ is called the partition function and is
simply given by the multiple integral

\begin{equation}
Z_N=\int\cdots \int \prod_i d\lambda_i \exp\left(- \mathcal{H}(\lambda_1,\dots,\lambda_N)
\right).
\label{partition}
\end{equation}

From \eqref{jpdeig}, one easily deduces that the system of $N$ eigenvalues of a classically invariant ensemble behaves as a thermodynamic system
of charged particles arranged on the real line, in equilibrium at inverse temperature $\beta$ under competing interactions (the confining potential $V(x)$ and the logarithmic all-to-all repulsion term)
in \eqref{ham}. In contrast with the usual canonical ensemble in statistical mechanics, however, the so-called Dyson index $\beta$ is quantized and can only assume the values $\beta=1,2,4$
for real symmetric, complex hermitian and quaternion self-dual matrices respectively. 

\end{enumerate} 

Lifting the quantization of $\beta$ (Dyson's \emph{threefold way}) has been a major 
theoretical challenge in view of possible applications e.g. to the quantum Hall effect. Dumitriu and Edelman \cite{Dumitriu} were eventually able to construct ensembles of 
tridiagonal matrices with independent entries whose eigenvalues are distributed as 
\eqref{jpdeig} with general $\beta >0$. Their ensemble is however \emph{not} invariant 
under similarity transformations, and the eigenvectors are not Haar distributed in the 
appropriate symmetry group. After an earlier attempt in the case of $2\times 2$ matrices 
\cite{Vivo}, the explicit construction of an ensemble of $N\times N$ matrices displaying 
at once rotational invariance and a continuous $\beta$ was put forward in 
\cite{GOE,paper_math} for the Gaussian ensemble. It was further shown in \cite{GOE} that 
only by letting the Dyson index $\beta$ of that ensemble scale with the matrix size $N$ in 
an appropriate way (namely $\beta=c/N$) one obtains a continuous family of deformed 
spectral densities parametrized by $c$, interpolating between Wigner's semicircle (typical 
for $\beta\sim\mathcal{O}(1)$ invariant ensembles) and a Gaussian law (properly describing 
the non-interacting limit $\beta\to 0$). This result can be established in two alternative 
ways:

\begin{enumerate}
\item Starting
from the stationary joint distribution of eigenvalues (eq. \eqref{jpdeig}), setting $\beta=c/N$,
and then finding the average density of eigenvalues $\rho(\lambda)=(1/N)\langle\sum_i\delta(\lambda-\lambda_i)\rangle$. 
In the limit of large $N$, this
average density can be obtained by a saddle point analysis of
the partition function Eq. \eqref{partition} in a standard way. Usually, when
$\beta\sim \mathcal{O}(1)$, only the energy term $\sim \mathcal{O}(N^2) $ dominates and the
entropy term $\sim \mathcal{O}(N) $ is subleading. However, when $\beta\sim c/N$, 
\emph{both} the
energy and the entropy terms are of the same order ($\sim \mathcal{O}(N)$), which
leads to a nontrivial modification of the density. The complicated nonlinear
singular integro-differential equation for the saddle density, reduces
very nicely to a Riccati equation for the Stieltjes transform of the
density, which can then subsequently solved exactly. Finally one
obtains the density by taking imaginary part of the Stieltjes transform
(see section \ref{subsection_saddle_p_meth} for a detailed discussion).

\item Starting from the dynamical equation of motion of the eigenvalues,
one first derives directly the equation of motion of the Stieltjes transform
of the density via It\^o's calculus, finds the stationary solution and 
then obtains the average density by taking the imaginary part
of the Stieltjes transform (see detailed discussion in section \ref{subsection_ito_m}).

In section \ref{subsection_saddle_p_meth} and \ref{subsection_ito_m}, in the context of our model, we show that both methods lead to the 
same solution.

\end{enumerate}

The purpose of this paper is threefold. We first explicitly construct a random matrix model $\iota)$ which is invariant under similarity transformations (and thus has Haar distributed eigenvectors), and $\iota\iota)$ whose jpd of eigenvalues is exactly given by the $\beta$-Wishart ensemble of random matrices with a continuous $\beta >0$. Then, letting the Dyson index $\beta$ of the ensemble scale 
inversely with the size of the matrix, we  analytically derive the density of states for this crossover model, 
written in terms of the Whittaker hypergeometric function (see eq. \eqref{sol_rho_crossover}), and we show that it continuously interpolates between the Mar\v cenko-Pastur law and 
a certain type of Gamma distribution (see subsection \ref{subcrossoversaddle}). 
Some other types of deformations of the Mar\v cenko-Pastur distribution for Wishart-like 
matrix models were 
reported in the literature (see e.g. \cite{AV,Burda}). Finally, keeping the Dyson index $\beta$ 
unscaled 
(i.e. it remains of $\sim\mathcal{O}(1)$ for large matrix size $N$) but nevertheless continuous ($\beta>0$), we analyze the full
Stieltjes transform equation and we can compute the $1/N$ correction to the Mar\v cenko-Pastur asymptotic density for the $\beta$-Wishart ensemble for all values of $\beta>0$. Furthermore, 
using results obtained by Pastur and Lytova in \cite{lytova} on the noise in the Mar\v cenko-Pastur law, we are also able to derive the $1/N^2$ correction term in the particular cases 
$\beta=1$ and $2$. 

The rest of the paper is outlined as follows. In section \ref{sec:Wishart} we introduce 
the main features of 
the classical Wishart ensemble along with the evolution law for the eigenvalue process.
In section \ref{diff_mat_beta}, we construct a $3$-parameters matrix model 
(and the respective evolution law for the eigenvalues) that at large times interpolates 
between
the Wishart ensemble and so-called CIR processes whose stationary pdf is a certain Gamma distribution. The corresponding parametrical density of states is computed exactly in the two ways described above (from the saddle point route on the partition function in section \ref{subsection_saddle_p_meth} and from It\^o's calculus in section \ref{subsection_ito_m})
and constitutes a continuous deformation of the Mar\v cenko-Pastur distribution (see below). 
In section \ref{section_cor_MP} we compute the systematic $1/N$ (for all $\beta>0$) and 
$1/N^2$ (for $\beta=1,2$) 
corrections to the Mar\v cenko-Pastur law for the smoothed density in the bulk for the 
(scaled) 
$\beta$-Wishart ensemble. For the special cases $\beta=1$, $2$ and $4$, the $\mathcal{O}(1/N)$
correction term was computed in ~\cite{FFG,Forrester12}. Our result generalizes this
to arbitrary $\beta$ for the $\mathcal{O}(1/N)$ term and in addition, we obtain the $\mathcal{O}(1/N^2)$
correction for $\beta=1$ and $\beta=2$. We 
conclude
with a summary and discussion in section \ref{concl}.

\section{Wishart ensembles}
\label{sec:Wishart}
\subsection{Real and complex Wishart ensembles}
Let ${\bf X}$ be a real (respectively complex) Gaussian random matrix of size $M\times N$, i.e. a random matrix chosen 
in the space of $M\times N$ real (resp. complex) matrices according to the law: 
\begin{equation}
P(\mathbf{X}) { d}\mathbf{X}\propto \exp\left(-\frac{1}{2\sigma^2} {\rm {Tr} }({\bf X}^\dagger {\bf X})\right) \,  { d}{\bf X}\,,
\end{equation}
where ${\bf X}^\dagger$ is the Hermitian conjugate of ${\bf X}$. In the following, we will denote the real (resp. complex) Wishart ensemble by $\mathcal{W}^\beta$ with $\beta=1$ in the real case (resp. $\beta=2$ in the complex case).

The real (resp. complex) Wishart Ensemble is the ensemble of ($N\times N$) square matrices 
of the product form ${\bf W}:={\bf X}^\dagger {\bf X}$ where ${\bf X}$ is a real (resp. 
complex) Gaussian random matrix of size $N\times M$. They have appeared in many different 
applications such as communication technology \cite{wishart1}, nuclear physics 
\cite{wishart2}, quantum chromodynamics \cite{wishart3}, statistical physics of directed 
polymers in random media \cite{wishart4} and non intersecting Brownian motions 
\cite{wishart5,wishart7,RS1}, as well as Principal Component Analysis of large datasets 
\cite{MV2009,wishart6}.

The spectral properties of the Wishart matrices have been studied extensively and it is known 
\cite{James} that for $M\geq N$, all $N$ positive 
eigenvalues of ${\bf W}$ are distributed 
via the joint probability density function (pdf) 
\begin{equation}
\label{P_beta}
P_\beta(\lambda_1,\dots,\lambda_N) = \frac{1}{Z} e^{-\frac{1}{2\sigma^2} \sum_{i=1}^N \lambda_i} \prod_{i=1}^N \lambda_i^{\frac{\beta}{2}(M-N+1)-1} 
\prod_{i < j} |\lambda_i - \lambda_j|^{\beta}
\end{equation}
where $Z$ is a constant normalization factor and where $\beta=1$ in the real case (resp. $\beta=2$ in the complex case). 
Note that the joint distribution $P_\beta$ defined in \eqref{P_beta} is in fact 
well behaved for every $\beta >0$.

Another classical result of Random Matrix Theory concerns the asymptotic density of states (or spectral measure) 
for the eigenvalues $(\lambda_1,\lambda_2,\dots,\lambda_N)$ 
of a real Wishart matrix ${\bf W}\in \mathcal{W}^1$  in the 
limit of large matrices, i.e. when $N,M \rightarrow \infty$ with $N/M=q\in (0;1]$ where $q$ is a fixed parameter. Let us recall that  
the density of states of the matrix ${\bf W}$ is simply the probability measure $\rho_N^\beta$ defined as 
\begin{equation}
\label{dsd_beta}
\rho_N^\beta = \frac{1}{N} \sum_{i=1}^N \delta(\lambda-\lambda_i)
\end{equation}
where $\beta$ is introduced for later convenience ($\beta\equiv 1$ in the present case) and
where $(\lambda_1,\dots,\lambda_N)$ are the eigenvalues of ${\bf W}$. Setting $\lambda=M \hat \lambda$, the Mar\v cenko-Pastur Theorem (see \cite{MP}) states that, in the limit $N,M\rightarrow \infty$ with 
$N/M=q\in (0;1]$, the spectral measure of a Wishart matrix ${\bf W}\in \mathcal{W}^1$ converges to a continuous probability density (with compact support) given by 
\begin{equation}
\label{marchenko}
\rho_\beta(\hat \lambda) = \frac{1}{2\pi \sigma^2 \beta q} \frac{\sqrt{(b-\hat \lambda)(\hat \lambda-a)}}{\hat\lambda} , \quad	a <\hat \lambda < b
\end{equation}
where the edges $a,b$ of the spectrum are given by 
\begin{equation*}
a= \sigma^2\beta \, (1-\sqrt{q})^2, \quad b= \sigma^2\beta \, (1+\sqrt{q})^2
\end{equation*}
with again $\beta=1$.

For general $\beta > 0$, the probability measure $\rho_N^\beta$ is defined 
again as in \eqref{dsd_beta} 
where this time the vector $(\lambda_1,\dots,\lambda_N)$ is 
distributed according to the law $P_\beta$ in \eqref{P_beta}.
The Mar\v cenko-Pastur theorem remains in fact valid for all $\beta>0$ in the sense that the probability law $\rho_N^\beta$ converges when $N,M\rightarrow \infty$ 
with $N/M=q \in (0;1]$ to the continuous probability density $\rho_\beta$ 
in \eqref{marchenko} for every $\beta >0$. 

The probability measure 
$\rho_N^\beta$ will sometimes be referred to as the spectral density as it corresponds to the spectral density of random matrices ${\bf W}\in \mathcal{W}^\beta$ 
at least when $\beta=1$ or $2$.

\subsection{Continuous processes for real and complex Wishart ensembles}
We wish to define here a diffusive matrix process depending on a fictitious time $t\geq 0$ that will converge to the Wishart Ensembles in the limit of large time. 
The idea is simply to set 
\begin{equation}
\label{def_W_t}
{\bf W}_t:= {\bf X}_t^\dagger {\bf X}_t
\end{equation}
where ${\bf X}_t$ is a real (resp. complex) random matrix process (of size $M\times N$) following the Ornstein-Uhlenbeck law,
\begin{equation*}
{ d}{\bf X}_t = -\frac{1}{2} {\bf X}_t { d}t + \sigma \, { d}{\bf B}_t
\end{equation*}
where ${\bf B}_t$ is a real Brownian (resp. complex) random matrix, i.e. 
a matrix whose entries are given by independent standard Brownian motions.
By a standard Brownian motion, one means a centered (zero-mean) Gaussian
process with covariance function $\langle B_t B_{t'}\rangle= {\rm min}(t,t')$. 

It is well known that the stationary law of a Ornstein-Uhlenbeck process is the Gaussian 
law and therefore, the real (resp. complex) matrix process ${\bf X}_t$ converges in law 
when $t\rightarrow \infty$ to the law of a Gaussian real (resp. complex) random matrix.  
Hence, we deduce that the real (resp. complex) matrix process ${\bf W}_t$ defines a 
diffusive matrix process that converges in law to ${\bf W}\in \mathcal{W}^\beta$
with $\beta=1$ or $\beta=2$.

It is also easy to check that the positive definite matrix 
process ${\bf W}_t$ verifies the following stochastic differential equation studied 
by Bru \cite{Bru}: 
\begin{equation}
{ d}{\bf W}_t = - {\bf W}_t  { d}t + \sigma \, \sqrt{{\bf W}_t} \,  { d}{\bf B}_t +  \sigma\,  { d}{\bf B}_t^\dagger \,  \sqrt{{\bf W}_t} + M\sigma^2\beta \, {\bf I} \, { d}t 
\end{equation}
where ${\bf B}_t$ is a real (resp. complex) Brownian random matrix and with $\beta=1$ in the real (resp. $\beta=2$ for complex) case.  

The evolution of the eigenvalue process $\lambda_1(t) \leq \lambda_2(t) \dots \leq \lambda_N(t)$ is also easy to derive \cite{Bru2} 
using perturbation theory to second order 
\begin{equation}
\label{SDE_ev}
{ d}\lambda_i = - \lambda_i { d}t + 2 \sigma \sqrt{\lambda_i}\, { d}b_i + 
\sigma^2  \beta \left(M + \sum_{k\neq i} \frac{\lambda_i+\lambda_k}{\lambda_i-\lambda_k} \right) { d}t
\end{equation}
where $b_i$'s are independent standard Brownian motions and with $\beta=1$ in the real 
(resp. $\beta=2$ for complex) case. 
The stationary distribution of the process $(\lambda_1,\dots,\lambda_N)(t)$ is 
necessarily the jpdf 
$P_\beta(\lambda_1,\dots,\lambda_N)$ defined in \eqref{P_beta} (this is true for any $\beta>0$ and can also be recovered using 
the Fokker-Planck equation for the multivariate diffusion \eqref{SDE_ev}).

\section{Crossover between Wishart and CIR processes}
\label{diff_mat_beta}

Following \cite{GOE,paper_math}, we aim at defining a diffusive matrix process ${\bf W}_t$ which converge in the limit of large time to a general $\beta$-Wishart matrix, i.e. a 
matrix whose eigenvalues 
are distributed according to $P_\beta$ in \eqref{P_beta} for general $\beta >0$ and with 
Haar distributed eigenvectors.  
In this paper, we will restrict ourselves to the description of the eigenvalues process but the interested reader can find a study of the eigenvectors for a related model in \cite{paper_math}. 
To simplify notations, we will take in this section $\sigma=1$. 

\subsection{Preliminary definition: CIR diffusion process}
\label{section_cir}

We first need to introduce a family of real diffusion processes. 
Let $\delta >0$ be a fixed parameter. The CIR process 
(named after its creators Cox, Ingersoll, and Ross~\cite{CIR85} and widely
used in finance to model short term interest rate) is a 
diffusion process 
$x(t)$ defined by $x(0):=x_0>0$ and for $t\geq 0$ by
\begin{equation}
\label{def_cir}
{ d}x(t) = -x(t) \, \,  { d}t +  2 \, \sqrt{x(t)} \, \, { d}b_t + \delta \, \, { d}t\,.
\end{equation} 
Using the assumption $\delta >0$, it is easy to see that the process $x(t)$ will remain non negative for all times $t\geq 0$. 
It is also easy to verify that the stationary pdf of the Langevin equation \eqref{def_cir} is the Gamma distribution with shape and scale parameters $k=\delta/2$ and $\theta=2$
defined as 
\begin{equation}
\label{CIR.dens}
p_\delta(x) = \frac{1}{2^{\frac{\delta}{2}} \Gamma(\frac{\delta}{2})}    
x^{\frac{\delta}{2}-1} e^{-\frac{x}{2}}  \,.
\end{equation}
In analogy with squared Bessel processes, the parameter $\delta$ will be called the \emph{dimension} of the process $x(t)$. 

\subsection{Diffusive matrix process for general $\beta$-Wishart matrices}\label{subsection_beta_wishart_process}
\label{diffmat}
Following \cite{GOE,paper_math}, our goal is to construct a diffusive matrix process whose eigenvalues process is asymptotically distributed according to 
$P_\beta$ for general $\beta\in [0,1]$. This construction can be extended respectively for general $\beta \in [0,2]$ (resp. $\beta\in [0,4]$) by using complex (resp. symplectic) 
Brownian motions instead of real Brownian motions in the following.  
 
We will in fact describe how to handle the value $\beta\in [0,1]$ by using real Brownian matrix. 
This construction can be extended for the values $\beta\in [0,2]$ using complex Brownian matrix and also $\beta\in [0,4]$ using symplectic 
Brownian matrix. 

The idea is to slice the time interval into small chops of length $1/n$ and for each interval $[k/n;(k+1)/n]$, to choose independently 
Bernoulli random variables $\epsilon_k^n, k\in\N$ such that $\P[\epsilon_k^n=1] = p = 1-\P[\epsilon_k^n=0]$. Then, setting $\epsilon_t^n = \epsilon_{[nt]}^n$, our diffusive matrix 
process evolves as:
\begin{equation}
\label{def_model}
{ d}{\bf W}_t^n = - {\bf W}_t^n \, { d}t + { d } {\bf \Delta}_t^n
\end{equation} 
where the increment matrix ${ d } {\bf \Delta}_t^n$ now depends on the value of the additional random process $\epsilon_t^n$:
\begin{itemize}
\item if $\epsilon_t^n=1$, then 
\begin{equation*}
{ d } {\bf \Delta}_t^n =
\sqrt{{\bf W}_t^n} \, { d}{\bf B}_t + { d}{\bf B}_t^\dagger \,  \sqrt{{\bf W}_t^n} +  M\,  \,{\bf I} \, { d}t.
\end{equation*}
where ${ d} {\bf B}_t$ is an $N\times N$ real \footnote{Here one can use use complex Brownian motions instead 
to extend the interval of $\beta$ to $[0,2]$.} Brownian increment matrix whose entries have variance ${ d}t$.  
\item  if $\epsilon_t^n=0$, then 
\begin{equation*}
{ d } {\bf \Delta}_t^n =
\sqrt{{\bf W}_t^n} \, { d}{\bf Y}_t + { d}{\bf Y}_t^\dagger \, \sqrt{{\bf W}_t^n} + \delta  \,  \,{\bf I} \, { d}t.
\end{equation*}
with $\delta>0$ and where ${ d}{\bf Y}_t$ is a symmetric matrix that is co-diagonalizable with ${\bf W}_t^n$ 
(i.e. the two matrix have the same eigenvectors)
but with a spectrum given by $N$ independent real Brownian increments of variance ${ d}t$.
\end{itemize}

An algorithmic description of how to build (approximatively on a discrete grid) the matrix process ${\bf W}_t^n$ can be found in Appendix \ref{algo}.  
 
It is clear that the eigenvalues of the matrix ${\bf W}_t ^n$ will cross at some points but only in intervals $[k/n;(k+1)/n]$ for which $\epsilon_k^n=0$ (in the other intervals 
where they 
follow the SDE \eqref{SDE_ev} with parameter $\beta=1$, it is well known that the repulsion is too strong and thus collisions are avoided). In this case, the eigenvalues are re-numbered 
at time $t=(k+1)/n$ in increasing order. With this procedure, when ordered $\lambda_1^n(t) \leq \dots \leq \lambda_N^n(t)$, 
we can again check as in \cite{GOE,paper_math}, using perturbation theory, that the eigenvalues will remain always non-negative and will verify the Stochastic Differential System (SDS): 
\begin{equation}
\label{sde_ev_n}
{ d}\lambda_i^n = - \lambda_i^n { d}t + 2  \sqrt{\lambda_i^n} \, { d}b_i +    \left( \epsilon_t^n M + (1-\epsilon_t^n) \delta  +  \epsilon_t^n  \sum_{k\neq i} 
\frac{\lambda_i^n+\lambda_k^n}{\lambda_i^n-\lambda_k^n} \right) { d}t
\end{equation}
where the $b_i$ are independent standard Brownian motions, which are also independent of the process $\epsilon_t^n$.

Note that when $\epsilon_t^n=0$, the particles $\lambda_i^n$ are evolving as independent CIR processes of 
dimension $\delta > 0$ as defined in paragraph \ref{section_cir}. Therefore, the particles can cross in those time intervals, breaking the 
increasing order so that they will be re-ordered at time $([nt] +1)/n$ but they {\it will} remain non-negative as the dimension $\delta$ is strictly positive. Therefore
the SDS \eqref{sde_ev_n} remains well defined at all times $t\geq 0$. 

One can follow the proof of \cite{paper_math} to prove that the scaling limit (i.e. the limiting process when $n\rightarrow \infty$) of the process $(\lambda_1^n(t)\leq\dots\leq \lambda_N^n(t))$ 
satisfies the following SDS 
\begin{equation}
\label{sde_ev_p}
{ d}\lambda_i = - \lambda_i { d}t + 2  \sqrt{\lambda_i} \, { d}b_i +    \left( p M + (1-p) \delta  +  p  \sum_{k\neq i} 
\frac{\lambda_i+\lambda_k}{\lambda_i-\lambda_k} \right) { d}t\,.
\end{equation}

One can deduce from the 
above equation \eqref{sde_ev_p} the Fokker-Planck equation for the joint density $P(\{\lambda_i\},t)$, for which the stationary jpdf is readily found to
be [see the derivation in appendix \ref{deriv_FP}]
\begin{equation}
\label{P_star}
P^*(\lambda_1,\dots,\lambda_N) = \frac{1}{Z} e^{-\frac{1}{2} \sum_{i=1}^N \lambda_i} 
\prod_{i=1}^N \lambda_i^{\frac{p}{2}(M-N+1-\delta)-(1-\frac{\delta}{2})} \prod_{i < j} |\lambda_i - \lambda_j|^{p}\,.
\end{equation}
The probability $P_\beta$ introduced in \eqref{P_beta} is recovered here by taking the values $p=\beta$ and $\delta=0$.   
The corresponding large $N,M$-limit spectral probability density is therefore given by the Mar\v cenko-Pastur law in the case where $p=\beta>0$ \emph{independent of $M$}. 
Note that with the above normalizations, the spectrum is spread over a region of $\R_+$ of width of order $pM =\beta M$.  On the other hand, if $p=0$, 
the large $N,M$-limit of the spectral density is the Gamma distribution with shape and scale parameters $k=\delta/2$ and 
$\theta=2$ (recall that it is the stationary pdf of the CIR process of dimension $\delta$): 
\begin{equation}
\rho_0(\lambda){ d} \lambda = \frac{1}{2^{\delta/2} \Gamma(\frac{\delta}{2})}    \lambda^{\frac{\delta}{2}-1} e^{-\frac{\lambda}{2}}  \, \, { d} \lambda\,. 
\end{equation}
It is quite natural to ask whether a crossover regime may be found, interpolating between the Mar\v cenko-Pastur density ($p$ independent of $M$) and the Gamma distribution ($p=0$). 
A good candidate for triggering such a transition is clearly a parameter $p$ vanishing with $M$ as $p=2c/M$ where $c$ is a positive fixed constant. We discuss this case in the following subsection.

In the next two subsections we compute the crossover density interpolating 
between the Mar\v cenko-Pastur 
law and the Gamma distribution with shape parameter $\delta/2$. This family of probability 
densities is indexed by the three parameters $c$ (such that $p=2c/M$), $q=N/M$ and 
$\delta\geq 0$. More precisely, we compute the limiting density of the probability 
measure $\rho_N= \frac{1}{N} \sum_{i=1}^N \delta_{\lambda_i}$, when $N,M\rightarrow \infty$ 
with $N/M=q\in (0;1]$ and where $(\lambda_1,\dots,\lambda_N)$ is distributed according to 
the law $P^*$ defined in \eqref{P_star} with $p = \beta = 2c/M$. As mentioned
in the introduction, the crossover density can be computed via two alternative 
methods: (1) by a saddle point method as shown in subsection \ref{subsection_saddle_p_meth} and (2) by
analyzing directly the stochastic differential systems introduced in \ref{subsection_beta_wishart_process} above
following the analogous route for the Gaussian case in Ref. ~\cite{GOE}. This is done in 
section \ref{subsection_ito_m}. We will see that both methods yield identical result.

\subsection{Crossover for the spectral density via the saddle point method}\label{subsection_saddle_p_meth}
\label{subcrossoversaddle}

Our starting point in the joint probability law of eigenvalues in Eq. (\ref{P_star}), 
where the normalization constant (partition function) $Z$ is given by the 
$N$-fold integral
\begin{align}
\nonumber Z &=\int_{[0,\infty]^N} \prod_i d\lambda_i e^{-\frac{1}{2}\sum_i \lambda_i}\prod_{i<j}|\lambda_i-\lambda_j|^p \prod_{i}\lambda_i^{\frac{p}{2}(M-N+1-\delta)-(1-\delta/2)}\\
&=\int_{[0,\infty]^N} \prod_i d\lambda_i e^{-E[\{\lambda_i\}]}
\label{partfun}
\end{align}
where the energy function $E[\{\lambda_i\}]$ is given by
\begin{equation}
E[\{\lambda_i\}]=\frac{1}{2}\sum_i \lambda_i -\left(\frac{p}{2}(M-N+1-\delta)-
(1-\delta/2)\right)\sum_i \ln \lambda_i-\frac{p}{2}\sum_{i\neq j}\ln 
|\lambda_i-\lambda_j|.
\label{energy.1}
\end{equation}
Written in this form, Eq. \eqref{partfun} is the Gibbs-Boltzmann canonical weight of a system of charged particles on the positive half-line in equilibrium at inverse temperature $\beta=1$ under
the effect of competing interactions. We wish to compute the
average density of states $\langle \frac{1}{N}\sum_{i=1}^N\delta(\lambda-\lambda_i)\rangle$ where
$\langle O \rangle$ denote the expectation value of $O$ with respect to
the probability distribution $P^*$ in \eqref{P_star}. There are many ways to compute this
average density, but the one rather convenient for large $N$ is the saddle point route.
This was originally done by Dyson~\cite{Dyson2} for the Gaussian random matrices and
a physically more transparent derivation can be found in Ref.~\cite{Satya-Dean}.

The main idea behind the saddle point 
method is as follows. In the large $N$ limit, the most dominant contribution
to the partition function emerges indeed from a set of configurations of $\lambda_i$'s
that correspond to a particular density $\rho^*(\lambda)$. Naturally then, the average
computed over the ensemble of $\lambda_i$'s, in this large $N$ limit, will also be given
by the saddle point density $\langle\frac{1}{N}\sum_{i=1}^N \delta(\lambda-\lambda_i)\rangle\approx 
\rho^*(\lambda)$. It thus suffices to analyze just the partition function $Z$
in the large $N$ limit and find, in particular, the saddle point density $\rho^*(\lambda)$
that maximizes the partition function $Z$ for large $N$.

To analyze $Z$ in the large $N$ limit, one first defines a `local' smooth density
function  
\begin{equation}
\rho(\lambda)=\frac{1}{N}\sum_{i=1}^N \delta\left(\lambda-\lambda_i\right)
\label{local_dens.1}
\end{equation}
which is normalized to unity. The main idea then is to split the multiple integration
in Eq. \eqref{partfun} in two parts: First fix the local density $\rho(\lambda)$ and sum 
over all microstates (i.e., configurations of $\lambda_i$'s consistent with 
the local density defined in \eqref{local_dens.1}) and then, sum (functionally)
over all possible local density functions. Roughly speaking, the first step corresponds to
a partial tracing over microstates by fixing the local density. Notationally, on can 
express 
this by
\begin{equation}
Z = \int {\cal D}[\rho] \int_{[0,\infty]^N} \prod_i d\lambda_i e^{-E[\{\lambda_i\}]}\, 
I\left[\rho(\lambda), \{\lambda_i\}\right]
\label{not.1}
\end{equation}
where ${\cal D}[\rho]$ denotes a functional integration over the function space
and $I\left[\rho(\lambda), \{\lambda_i\}\right]$ is an indicator function
that is $1$ if the microstate $\{\lambda_i\}$ is compatible with a given $\rho(\lambda)$,
normalized to unity, as defined in \eqref{local_dens.1} and otherwise $I=0$.
The energy function $E[\{\lambda_i\}]$ associated with a microstate can then
be expressed in terms of the local density $\rho(\lambda)$        
using the identity $\sum_i f(\lambda_i)=N\int d\lambda f(\lambda)\rho(\lambda)$ 
and one gets~\cite{Satya-Dean}
\begin{align}
\nonumber E[\rho(\lambda)] &= \frac{N}{2}\int d\lambda \lambda \rho(\lambda) -\left[\frac{p}{2}\left((\frac{1}{q}-1)N+1-\delta \right)-\left(1-\frac{\delta}{2}\right)\right] N\int d\lambda\rho(\lambda)\ln \lambda\\
& -\frac{p}{2}N^2 \int\int d\lambda d\lambda^\prime \rho(\lambda)\rho(\lambda^\prime)\ln |\lambda-\lambda^\prime|+\frac{p}{2}N\int d\lambda \rho(\lambda)\ln \frac{1}{\rho(\lambda)}
+C_1\left(\int d\lambda \rho(\lambda)-1\right)
\end{align}
where the last term includes a Lagrange multiplier $C_1$ that enforces the 
normalization of the local density to $1$. The next-to-last term accounts for the 
self-energy term $(\lambda\to\lambda^\prime)$ 
that needs to be subtracted. Note indeed that in the original discrete sum $\sum_{i\neq j}\ln |\lambda_i-\lambda_j|$,
the eigenvalues do not coincide. This means that the integral over 
$\lambda$ and $\lambda'$ should exclude the region where $|\lambda - \lambda'|$ is less than the typical spacing between 
eigenvalues, which is proportional to $1/N \rho(\lambda)$~\cite{Dyson2}. The contribution 
of this thin 
sliver is the next-to-last term, up to an additional contribution that can be absorbed into 
$C_1$. 

Once this is done, equation \eqref{not.1} simplifies further and one gets
\begin{equation}
Z \approx \int {\cal D}[\rho]\, e^{-E[\rho(\lambda)]}\, 
J[\rho(\lambda)]
\label{not.2}
\end{equation}
where $J[\rho(\lambda)]= \int_{[0,\infty]^N} \prod_i d\lambda_i I\left[\rho(\lambda), 
\{\lambda_i\}\right]$ is an entropic factor that just counts how many microstates
are compatible with a given local density function $\rho(\lambda)$. This can
be estimated very simply by the following combinatorial argument involving
arrangement of $N$ particles in $K$ boxes. Let us first divide our one dimensional
line into $K$ small boxes of equal width. We have $N$ particles that need to be
distributed into the $K$ boxes with occupation numbers $\{n_1,n_2,\ldots, n_K\}$.
The number of ways this can be done is simply    
\begin{equation}
\frac{N!}{n_1! n_2! \cdots n_K!} \ .
\end{equation}
Setting $\rho_i = n_i/N$ (the local density in box $i$) and using Stirling's 
approximation $N!\sim N^{N+1/2}\,e^{-N}$ (using the fact that $\sum_i n_i=N$), we have:
\begin{equation}
\frac{N!}{n_1! n_2! \cdots n_K!}\sim e^{-\sum_i n_i \ln n_i}
\end{equation}
which in the continuum limit, becomes 
$\sim e^{-N\int d\lambda \rho(\lambda)\ln \rho(\lambda)}$. Thus, the entropic 
factor can also be expressed as a simple functional of the local density $\rho(\lambda)$. 
Inserting this expression in the functional integral over the density, yields:
\begin{equation}
Z = \int \mathcal{D}[\rho]e^{-E[\rho(\lambda)]}e^{-N\int d\lambda \rho(\lambda)\ln \rho(\lambda)}=\int \mathcal{D}[\rho] e^{-N F[\rho(\lambda)]}
\end{equation}
where the free energy $F[\rho(\lambda)]$ is given by:
\begin{align}
\nonumber F &[\rho(\lambda)] = \frac{1}{2}\int d\lambda \lambda \rho(\lambda) -
\left[\frac{p}{2}\left((\frac{1}{q}-1)N+1-\delta \right)-\left(1-\frac{\delta}{2}\right)\right] \int d\lambda\rho(\lambda)\ln \lambda\\
& -\frac{p}{2}N \int\int d\lambda d\lambda^\prime \rho(\lambda)\rho(\lambda^\prime)\ln |\lambda-\lambda^\prime|+\left(1-\frac{p}{2}\right)\int d\lambda \rho(\lambda)\ln \rho(\lambda)
+C_1\left(\int d\lambda \rho(\lambda)-1\right)
\end{align}

Note that for $p\sim\mathcal{O}(1/N)$ the entropy term becomes of the same order of the energy term, while in the usual case $p\sim\mathcal{O}(1)$
the entropy contribution is subdominant in the large $N$ limit and is therefore disregarded.

Setting now $p=2 c/M=2 c q/N$, we get:

\begin{align}
\nonumber F &[\rho(\lambda)] = 
\frac{1}{2}\int d\lambda \lambda \rho(\lambda) -
\left[cq\left(\frac{1}{q}-1\right)-\left(1-\frac{\delta}{2}\right)\right] \int d\lambda\rho(\lambda)\ln \lambda\\
& -c q \int\int d\lambda d\lambda^\prime \rho(\lambda)\rho(\lambda^\prime)
\ln |\lambda-\lambda^\prime|+\left(1-\frac{cq}{N}\right)\int d\lambda \rho(\lambda)\ln 
\rho(\lambda)
+C_1\left(\int d\lambda \rho(\lambda)-1\right)
\end{align}

We set $a=cq (1/q-1)-(1-\delta/2)$ and take $N\to \infty$ (so that
the term $cq/N$ drops out). Now, the saddle point density $\rho^*(\lambda)$
is obtained by minimizing the free energy $F[\rho(\lambda)$, i.e., by
taking the functional derivative 
$\frac{\delta F}{\delta \rho}=0$ and $\rho(\lambda)==\rho^*(\lambda)$.
This gives the saddle point equation
\begin{equation}
\label{sadd.1}
\frac{\lambda}{2}-a \ln\lambda -2c q\int d\lambda^\prime \rho^*(\lambda^\prime)\ln 
|\lambda-\lambda^\prime|+\ln\rho^* + C_2=0 
\end{equation}
where $C_2=C_1+1$ is just a constant. 
For notational simplicity, in the rest of the subsection we will denote the saddle point 
density $\rho^*(\lambda)$ simply by $\rho(\lambda)$.

Taking one more derivative of \eqref{sadd.1}, we get
\begin{equation}
\frac{1}{2}-\frac{a}{\lambda}-2 c q\ 
\mathrm{Pr}\int\frac{\rho(\lambda^\prime)}{\lambda-\lambda^\prime}d\lambda^\prime+
\frac{\rho^\prime 
(\lambda)}{\rho(\lambda)}=0
\label{fund} 
\end{equation}

Next, we define the Stieltjes transform:
\begin{equation}
H(z)=\int\frac{\rho(\lambda)}{\lambda-z}d\lambda
\end{equation}
for $z$ complex and outside the support of $\rho$. By definition, for large $|z|$, $H(z)\to -1/z$. Multiplying eq. \eqref{fund} by $\rho(\lambda)/(\lambda-z)$ and integrating over $\lambda$, we have:
\begin{equation}
\frac{1}{2}\int \frac{\rho(\lambda)}{\lambda-z}d\lambda-a\int \frac{\rho(\lambda)d\lambda}{\lambda (\lambda-z)}-2 c q\ \mathrm{Pr}\int \frac{\rho(\lambda)d\lambda}{\lambda-z}\int\frac{\rho(\lambda^\prime)}{\lambda-\lambda^\prime}d\lambda^\prime+\int \frac{\rho^\prime (\lambda) d\lambda}{\lambda-z}=0\label{fund2}
\end{equation}

and we analyze each of the four contributions separately.
\begin{enumerate}
\item $T_1=\frac{1}{2}\int \frac{\rho(\lambda)}{\lambda-z}d\lambda=\frac{1}{2}H(z)$
\item $T_2 =-a\int \frac{\rho(\lambda)d\lambda}{\lambda (\lambda-z)} $. We rewrite this as:
\begin{equation}
T_2 =-a\int \rho(\lambda)d\lambda \left[\frac{1}{\lambda-z}-\frac{1}{\lambda}\right]\frac{1}{z}
\end{equation}
implying:
\begin{equation}
T_2=-\frac{a}{z}H(z)+\frac{b_1}{z}
\end{equation}
where $b_1=a\int d\lambda\frac{\rho(\lambda)}{\lambda}$.
\item $T_3 = -2 c q\ \mathrm{Pr}\int \frac{\rho(\lambda)d\lambda}{\lambda-z}\int\frac{\rho(\lambda^\prime)}{\lambda-\lambda^\prime}d\lambda^\prime$
which we rewrite as:
\begin{align}
T_3 &=2c q \left\{\mathrm{Pr}\int d\lambda d\lambda^\prime \rho(\lambda)\rho(\lambda^\prime)\left[\frac{1}{\lambda-z}-\frac{1}{\lambda-\lambda^\prime}\right]\frac{1}{\lambda^\prime-z}\right\}\\
&= 2c q H^2(z)-2c q\ \mathrm{Pr}\int\frac{d\lambda d\lambda^\prime \rho(\lambda)\rho(\lambda^\prime)}{(\lambda-\lambda^\prime)(\lambda^\prime-z)}
\end{align}
By renaming $\lambda\to\lambda^\prime$ and $\lambda^\prime\to\lambda$, we get:
\begin{align}
T_3 &= 2 c q H^2(z)+2c q\mathrm{Pr}\int\frac{d\lambda d\lambda^\prime \rho(\lambda)\rho(\lambda^\prime)}{(\lambda-\lambda^\prime)(\lambda-z)}\\
&=2c q H^2(z)-T_3
\end{align}
Solving for $T_3$ we get:
\begin{equation}
T_3=c q H^2(z)
\end{equation}

\item $T_4=\int \frac{\rho^\prime (\lambda) d\lambda}{\lambda-z}$, which we integrate by parts, obtaining:

\begin{align}
\nonumber T_4 &=\frac{1}{\lambda-z}\rho(\lambda)\Big|_0^\infty+\int \frac{\rho(\lambda)}{(\lambda-z)^2}d\lambda\\
&=\frac{c_1}{z}+H^\prime (z)
\end{align}

\end{enumerate}

In the derivation above, we assumed $b_1$ and $c_1$ to be finite. This is not completely obvious, because $\rho(\lambda)$ at an edge point
may diverge. However, by imposing that for large $z$, $H(z)\to -1/z$, it is immediate to derive that $b_1+c_1=1/2$.
Thus, one may regularize the density near the edge points so that
$b_1$ and $c_1$ exist individually, but eventually
their sum is universally $1/2$
and hence is independent of the specific regularization near the edge.  

Adding up the four contributions, we get the equation:
\begin{equation*}
\frac{{ d}H}{{ d}z}  -\frac{a}{z}H +  \frac{b_1+c_1}{z}+c q \, H^2 + \frac{1}{2}H =0
\end{equation*}
Thus we find the following differential equation for the Stieltjes transform $H$:
\begin{equation}
\label{eq_H_crossover}
\boxed{\frac{{ d}H}{{ d}z} + \gamma \, H^2 +\frac{1}{2}\left(1+\frac{\alpha}{z}\right)\, H+ \frac{1}{2z}=0}
\end{equation}
where we have set  
\begin{equation*}
\alpha =  (2-\delta)- 2c(1-q) ,\quad \gamma = c q \,. 
\end{equation*} 
In the next subsection, we will derive the same equation via It\^o's stochastic calculus route.

The density $\rho(\lambda)$ (normalized to unity)
can then be read off from
\begin{equation}
\rho(\lambda)= \frac{1}{\pi} {\rm Im} [H(z\to \lambda)]
\label{dens.1}
\end{equation}
where $z\to \lambda$ occurs inside the cut on the real axis. 

To solve the Riccati equation \eqref{eq_H_crossover}, we make a standard 
substitution 
\begin{equation}
H(z) = \frac{1}{\gamma} \frac{u'(z)}{u(z)}= \frac{1}{\gamma}\,\partial_z \ln u(z)\, .
\label{cole.1}
\end{equation}
This gives a second order differential equation for $u(z)$
\begin{equation}\label{eq_u_z}
u''(z) + \frac{1}{2} \left[1+\frac{\alpha}{z}\right] u'(z) + \frac{\gamma }{2 z} u(z) = 0\,.
\end{equation}
It follows from Eq. (\ref{cole.1}) and the asymptotic behavior of $H(z)$ that
\begin{equation}
u(z)\xrightarrow[|z|\to \infty]{} \frac{A_1}{z^{\gamma}}
\label{uzasymp}
\end{equation}
where $A_1$ is a constant.

To reduce Eq. \eqref{eq_u_z} to a Schr\"odinger like differential equation, we make the substitution 
\begin{equation}
u(z) = e^{-z/4} \, z^{\alpha/4} \, \psi(z) \, ,
\label{uz.4}
\end{equation}
and we find the following equation for $\psi$
\begin{equation*}
\psi''(z) + \left[ - \frac{1}{16} + \frac{1}{z} \frac{4 \gamma - \alpha}{8} + 
\frac{\alpha}{4} \left(1-\frac{\alpha}{4}\right)\,  \frac{1}{z^2}  \right] \psi(z) = 0\,.
\end{equation*}
Making further a rescaling $\psi(z) = y(z/2)$, it reduces to the standard form of the Whittaker 
differential equation~\cite{GR}
\begin{equation}\label{Whitt.1}
y''(z) + \left[ -\frac{1}{4} + \frac{\lambda}{z} + \frac{\frac{1}{4}- \mu^2}{z^2}  \right] y(z) = 0\,,
\end{equation}
where 
\begin{equation}\label{param.1}
\zeta = \gamma - \frac{\alpha}{4}, \quad \mu = \frac{1}{2} |1-\frac{\alpha}{2}|\,.
\end{equation}
Note that the solution of this differential equation does not depend on the sign of $\mu$, 
hence we take the absolute value. The differential equation (\ref{Whitt.1}) has two linearly 
independent solutions $W_{\zeta,\mu}(z)$ and $W_{-\zeta,\mu}(-z)$. The Whittaker
function $W_{\zeta,\mu}(z)$ has the following asymptotic behavior~\cite{GR}
\begin{equation}
W_{\zeta,\mu}(z)\xrightarrow[|z|\to \infty]{} z^{\zeta}\, e^{-z/2}\,.
\label{Whitt.asymp}
\end{equation}
Thus the general solution of $u(z)$, using Eq. (\ref{uz.4}), reads
\begin{equation}
u(z) =  e^{-z/4} \, z^{\alpha/4} \, \left[{\cal C}_1\, W_{\zeta,\mu}(z/2)+ {\cal C}_2\, 
W_{-\zeta,\mu}(-z/2)\right]
\label{sol.2}
\end{equation}
where ${\cal C}_1$ and ${\cal C}_2$ are arbitrary constants.
Using the asymptotic behavior in Eq. (\ref{Whitt.asymp}) it is easy to check
that only the second solution has the right asymptotic behavior in Eq. (\ref{uzasymp}).
Thus, finally, we have our solution
\begin{equation}
u(z) = {\cal C}_2\, e^{-z/4} \, z^{\alpha/4} \, W_{-\zeta,\mu}(-z/2)
\label{sol.3}
\end{equation}
where $\zeta$ and $\mu$ are given in Eq. (\ref{param.1}).

By plugging this solution \eqref{sol.3} into Eq. \eqref{cole.1} and using Eq. \eqref{dens.1}, we find the following expression
\begin{equation}
\label{dens.2}
\rho(\lambda) = \frac{{\cal C}_2}{2 \pi \gamma } \frac{({\rm Im}(W'_{-\zeta,\mu}) 
{\rm Re}(W_{-\zeta,\mu}) - {\rm Im}(W_{-\zeta,\mu}) 
{\rm Re}(W'_{-\zeta,\mu}))(-\lambda/2)}{|W_{-\zeta,\mu}(-\lambda/2)|^2} \,. 
\end{equation}
where ${\rm Re}$ and ${\rm Im}$ denote respectively the real and imaginary parts.
Using the linear differential equation verified by the Whittaker functions 
\eqref{Whitt.1}, it is easy to see that the derivative with respect to 
$\lambda$ of the Wronskian type 
function $({\rm Im}(W'_{-\zeta,\mu}) {\rm Re}(W_{-\zeta,\mu}) - {\rm Im}(W_{-\zeta,\mu}) 
{\rm Re}(W'_{-\zeta,\mu}))$ is equal to $0$. Hence the Wronskian appearing
in the numerator in \eqref{dens.2} is simply a constant.  

Collecting all the constants together, we get:
\begin{equation}
\rho(\lambda)= \frac{A}{|W_{-\zeta,\mu}(-\lambda/2)|^2}\, .
\label{sol.4}
\end{equation}
The overall normalization constant $A$ has to be fixed from $\int_0^{\infty} \rho(\lambda)\,{ d}\lambda=1$.
Thus we get, after rescaling $\lambda/2\to \lambda$,
\begin{equation}
\frac{1}{A}= 2 \int_0^{\infty} \frac{{ d}\lambda}{|W_{-\zeta,\mu}(-\lambda)|^2}\,.
\label{norm.1}
\end{equation}
This integral in Eq. (\ref{norm.1}) can be done in closed form. First, we first use the well known identity~\cite{AS}
\begin{equation}
W_{\zeta,\mu}(z)= z^{\mu+1/2}\, e^{-z/2}\, U(\mu-\zeta+1/2,1+2\mu; z)
\label{iden.1}
\end{equation}
where $U(a,b;z)$ is the Tricomi hypergeometric function (or Kummer function) that behaves 
for large $z$
as $U(z)\sim z^{-a}$. Using this in Eq. (\ref{norm.1}) gives
\begin{equation}
\frac{1}{A}= 2 \, \int_0^{\infty}\, { d} \lambda \, \lambda^{-2\mu-1}\, e^{-\lambda}\, |U(\mu+\zeta+1/2,1+2\mu; -\lambda)|^{-2}\,.
\label{norm.2}
\end{equation}
It turns out that there exists an interesting integral representation in a paper by Ismail and 
Kelker~\cite{IK}
\begin{equation}
\int_0^{\infty}\frac{ dt\ e^{-t}\, t^{-b}}{z+t}\, |U(a,b; -t)|^{-2}
=\Gamma(a)\Gamma(a-b+2)\, \frac{1}{z}\frac{U(a,b-1;z)}{U(a,b;z)};\quad {\rm for}\, a>0,\, 
1<b<a+1
\label{iden.2}
\end{equation}
Note that in Ref. \cite{IK} they use the notation $\psi(a,b,z)$ instead of $U(a,b;z)$, but
it is the same function.  Our $\mu$ and $\zeta$ satisfy the condition
of validity of this identity: $a>0$ and $1<b<a+1$. Taking $z\to \infty$ limit on both sides and using $U(z)\sim z^{-a}$, we arrive
at the following exact expression of the normalization constant
\begin{equation}
\frac{1}{A}= 2 \, \Gamma(\mu+\zeta+1/2)\Gamma(\zeta-\mu+3/2)\,.
\label{norm.3}
\end{equation}

This leads to the following final expression for the spectral density, which is the central result of our work\footnote{We add the subscript $c$ in the notation of the density $\rho$ to recall the dependence in $c$; the density $\rho$ depends also on the two parameters $\delta$ and $q$ but we omit to subscript those.}:
\begin{equation}\label{sol_rho_crossover}
\boxed{\rho_c(\lambda) = \frac{1}{2 \Gamma(\mu+\zeta+\frac{1}{2}) 
\Gamma(\zeta-\mu+\frac{3}{2})} \frac{1}{|W_{-\zeta,\mu}(-\frac{\lambda}{2})|^2} }
\end{equation}
with the following values for the parameters 
\begin{equation*}
 \alpha= (2-\delta)- 2c(1-q);\quad \zeta = cq -\frac{\alpha}{4}; \quad {\rm and} \quad \mu=\frac{1}{4}|\alpha-2|\, .
\end{equation*}
The above expression is the analogue, in the present context, of the Askey-Wimp-Kerov one-parameter family of models found in \cite{GOE}, that smoothly interpolates between 
the Gaussian distribution and Wigner's semi-circle.

Let us now consider the limiting case $c\to 0$ first. In this case, we have 
$\alpha=(2-\delta)$, $\zeta=\delta/4-1/2$ and $\mu=\delta/4$. Thus,
$W_{-\zeta,\mu}(-\lambda/2)= W_{1/2-\delta/4,\delta/4}(-\lambda/2)$.
It turns out that for these special values of the indices, the Whittaker function
simply reduces to $W_{1/2-\delta/4,\delta/4}(-\lambda/2) \propto \lambda^{-\delta/4+1/2}\,e^{\lambda/4}$
up to a proportionality constant~\cite{GR}. Substituting this in 
\eqref{sol_rho_crossover},
we then recover the CIR density in Eq. \eqref{CIR.dens}. 
The limit $c\to \infty$ is more 
tricky
as one needs to rescale $\lambda\to \hat \lambda \, c$ and take the large $c$ limit
carefully. This can be done and one recovers the Mar\v cenko-Pastur law. 
This can be rewritten (without rescaling $\lambda$) as 
\begin{equation}\label{MP_c_infty}
\rho_c(\lambda) \sim_{c\to\infty} \frac{1}{4\pi c q} \frac{\sqrt{(\gamma_+-\lambda)(\lambda-\gamma_-)}}{\lambda} \, \, {\mathbf 1}_ {\{\gamma_- < \lambda < \gamma_+\}} \,,  
\end{equation}
where $\gamma_\pm=2c \, (1\pm\sqrt{q})^2$.

Using standard results on Whittaker functions (see e.g. \cite{GR,AS}), it is easy to compute the asymptotic behavior of $\rho_c(\lambda)$ for $\lambda\rightarrow 0$ and $\lambda\rightarrow +\infty$.
Up to multiplicative constants, we have
\begin{equation*}
\rho_c(\lambda) \sim_{\lambda\rightarrow +\infty} \lambda^{2\zeta} \, \, e^{-\frac{\lambda}{2}}\,,
\end{equation*}
and 
\begin{equation*}
\rho_c(\lambda) \sim_{\lambda\rightarrow 0_+} \lambda^{2\mu-1} \,.
\end{equation*}

\begin{figure}[h!btp] 
     \center
     \includegraphics[scale=0.75]{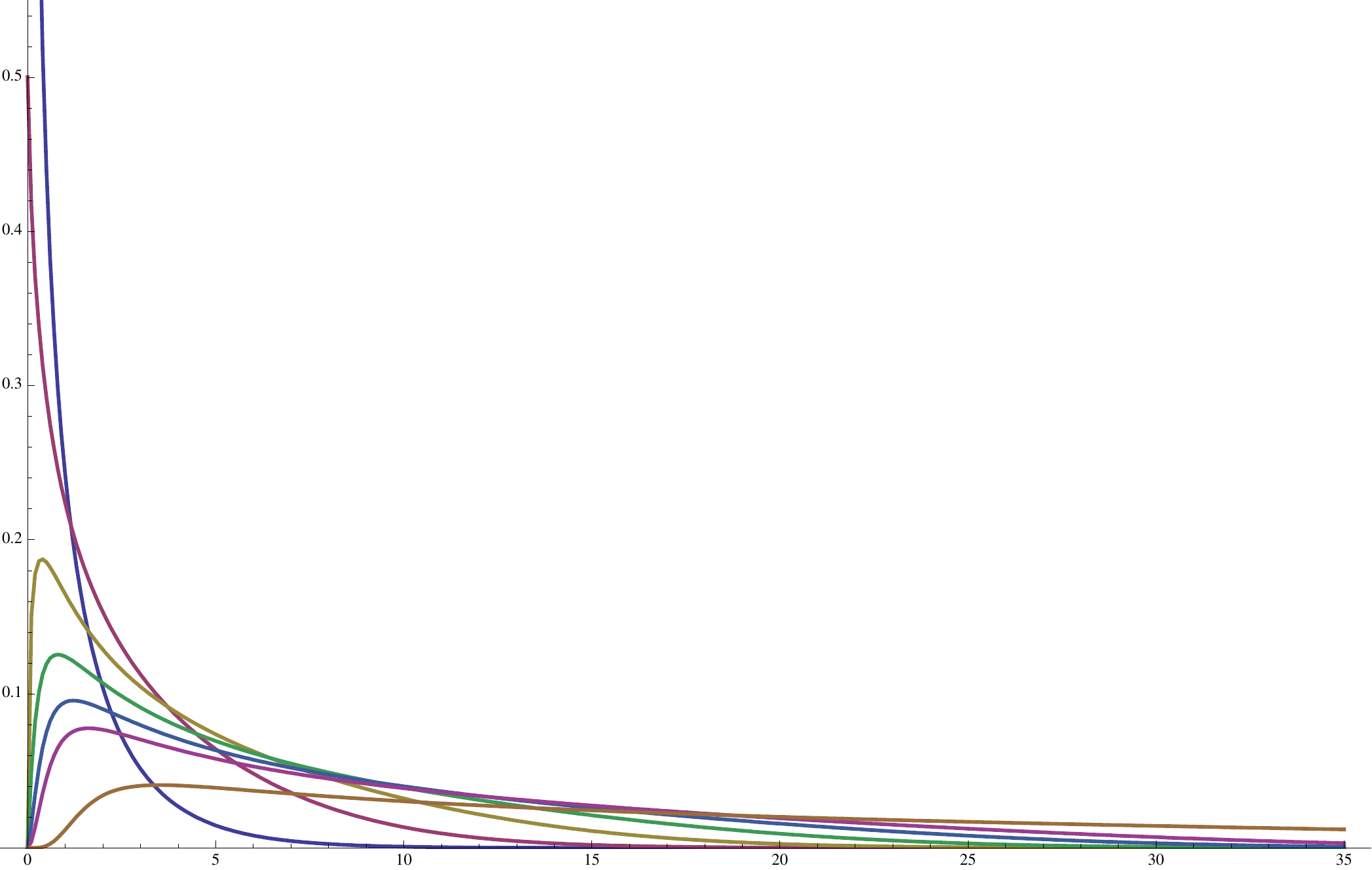}
     \caption{Density $\rho_c(\lambda)$ for $c=0,1,2,3,4,5,10$ of Eq. \eqref{sol_rho_crossover}  showing the progressive deformation of the Gamma distribution \eqref{CIR.dens} with parameter $\delta=1$ towards
     the Mar\v cenko-Pastur distribution with parameter $q=1/2$. The value $\rho_c(0)$ at the origin decreases when $c$ increases. }\label{fig_1}
\end{figure}

We plotted in Fig. \ref{fig_1} the density $\rho_c$ for $c=0,1,2,3,4,5,10$ and $q=1/2,\delta=1$, showing the progressive deformation 
of the Gamma distribution with shape parameter $\delta/2$ towards the Mar\v cenko-Pastur distribution \eqref{MP_c_infty} with parameter $q=1/2$.  
The critical value of $c$ at which the divergence at $\lambda \rightarrow 0_+$ changes to convergence is $c^*= (2-\delta)/(2(1-q))=1$. 
As expected, in Fig. \ref{fig_1}, the curve with second highest value at the origin corresponds to $c=1$ and converges when $\lambda\rightarrow 0_+$ to $1/2$.  
The curve with highest value at the origin is the Gamma distribution with shape parameter $\delta/2=1/2$ and diverges at $0_+$ to $+\infty$. The other curves corresponding to $c=2,3,4,5,10$ 
converge to $0$ when $\lambda\rightarrow 0_+$. 

\begin{figure}[h!btp] 
	\center
	\includegraphics[scale=0.8]{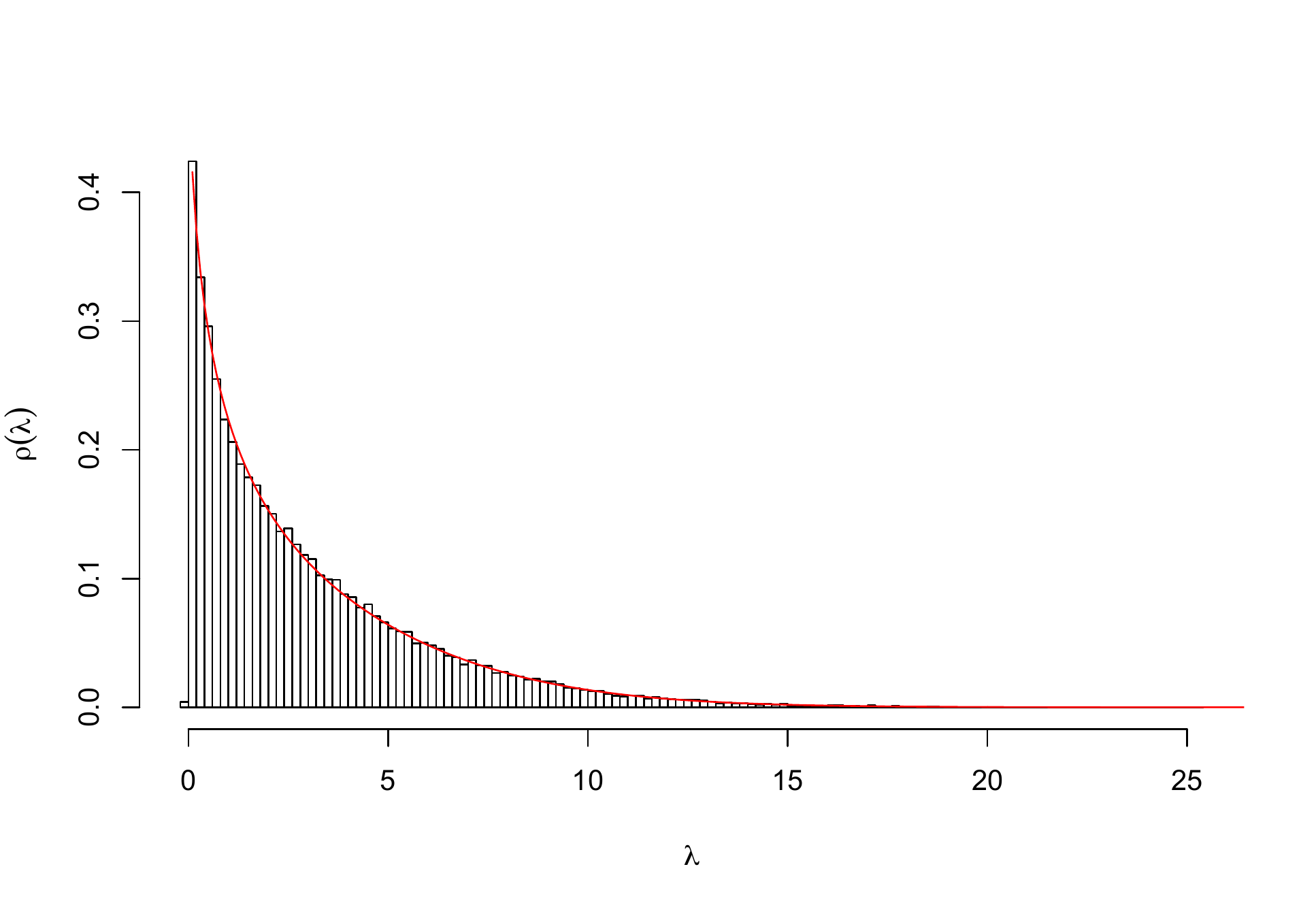}
	\caption{Numerical simulations of the state density of the random matrix ${\bf W}_{t=\infty}^n$ defined in Eq. \eqref{def_model} for $c=1,p=\beta=2c/M,M=100,N=50,\delta=1,q=1/2$.  }\label{fig_2}
\end{figure}

We have also verified our analytical result for the crossover density in 
\eqref{sol_rho_crossover} 
numerically for the sample value $c=1$
and found very good agreement (see Fig. \ref{fig_2}).  

\subsection{Crossover for the spectral density via It\^o's stochastic calculus}\label{subsection_ito_m}
\label{subcrossoverIto}
In this subsection, we want to re-obtain the result Eq. \eqref{eq_H_crossover} of the previous subsection via It\^o's calculus.  
We therefore consider the process $(\lambda_1(t),\cdots,\lambda_N(t))$ which verifies the stochastic differential system \eqref{sde_ev_p} with the scaling 
relation $p=2c/M$. The idea is to work out 
the evolution equation of the probability measure 
\begin{equation}
\label{emp_meas_lambda_t}
\rho_N^t(dx) := \frac{1}{N} \sum_{i=1}^N \delta(x-\lambda_i(t))
\end{equation} 
in the large $N$ limit. We expect the equilibrium 
of this evolution equation to be the solution of \eqref{eq_H_crossover}.  

In the following, $f$ is a smooth function. 
Using It\^o's formula for $\int f(x) \rho_N^t( dx)$, Eq. \eqref{sde_ev_p} and the scaling relation $p=2c/M$, we obtain (see \cite{Rogers_Shi} for similar calculations)
\begin{align}
\label{evolution_mu_t}
d \int f(x) \rho_N^t(dx) &=  \int \left(- x + 2c + \left(1-\frac{2c}{M}\right)\delta \right) f'(x) \rho_N^s(dx) dt   \\
 &+ 2\left(1-\frac{c}{M}\right) \int x f''(x) \rho_N^s(dx) dt \notag\\
 &+ cq \int_0^t \int \int \frac{f'(x)-f'(y)}{x-y} (x+y) \rho_N^s(dx) \rho_N^s(dy) dt +   dM_t^N \notag   \,,
\end{align}
where $dM_t^N = \frac{2}{N} \sum_{i=1}^N \sqrt{\lambda_i} \, f'(\lambda_i) db_i$ is a noise term of variance 
$\frac{4}{N} \int \lambda f'(\lambda)^2\rho_N^t({ d}\lambda) { d}t$. When $N,M\rightarrow \infty$ with $N/M=q$, this noise term is of order $1/\sqrt{N}$.

In the large $N,M$ limit, the {\it stationary} probability measure $\rho$ 
solution of Eq. \eqref{evolution_mu_t} therefore satisfies to leading order (keeping only the terms of order $1$)
\begin{align}
\label{eq_with_f}
 \int \left(- x + 2c + \delta \right) f'(x) \rho(dx) +  2 \int x f''(x) \rho(dx) \\
 + cq \int \int \frac{f'(x)-f'(y)}{x-y} (x+y) \rho(dx) \rho(dy) =0 \notag
\end{align}
Applying Eq. \eqref{eq_with_f} to the particular function $f(x)= \frac{1}{x-z}$ for $z\in \C\setminus\R$ and denoting $H(z)$ the Stieltjes transform of the probability measure $\rho$, we 
obtain the following differential equation for $H$
 \begin{align}
\label{eq_diff_H_ito_1}
\left[H(z)+ z H'(z)\right] -  \left(2c +\delta \right) H'(z) +  2 \left[2 H'(z)+ z H''(z)\right] \notag\\
+ 2cq \left[H'(z) + H^2(z) + 2 z H(z) H'(z) \right]= 0\,. 
\end{align}
Eq. \eqref{eq_diff_H_ito_1} can be rearranged as
\begin{align}
\label{eq_diff_H_ito_2}
2cq H(z) [H(z) + 2 z  H'(z)] + \frac{1}{2} [H(z)+2zH'(z)] + \frac{1}{2} H(z) 
+ [3 H'(z) + 2z H''(z)]\notag \\ + \left[(1-\delta)- 2c(1-q)\right] H'(z) = 0\,.
\end{align}
Eq. \eqref{eq_diff_H_ito_2} can be integrated easily by doing the change of function $G(z) = z H(z^2)$. 
Indeed, we just need to write \eqref{eq_diff_H_ito_2} for $z^2$ instead of $z$ and then multiply the corresponding equation by $z$ to obtain the following equation
\begin{align*}
2cq G(z) G'(z) +  \frac{1}{2}[z G'(z) + G(z)] +  \frac{1}{2} G''(z) + \left[\frac{1-\delta}{2} - c(1-q)\right] 2z H'(z^2) = 0\,,
\end{align*}
which can be integrated as
\begin{align}
\label{eq_G_ito}
2cq G^2(z)  +  \left[ z +  \frac{(1-\delta)-2c(1-q)}{z} \right] G(z) +  G'(z)  = -1 
\end{align}
where the integration constant is chosen so that $z G(z) \sim -1$ when $|z|\rightarrow \infty$. 
Note that the asymptotic behavior for $H$ is therefore also $zH\sim -1$ when $|z|\rightarrow \infty$ as is expected for the Stieltjes transform of a probability measure. 
Rewriting now Equation \eqref{eq_G_ito} in term of the function $H$, we obtain exactly Eq. \eqref{eq_H_crossover}.

\section{Correction to the Mar\v cenko-Pastur law for large but finite dimension}
\label{section_cor_MP}

In this section, we come back to the case of generalized Wishart matrices for which 
particles are distributed according to the jpdf $P_\beta$ with general parameter $\beta>0$ 
(\emph{not} scaling with $M$). We want to compute the first correction terms to the Mar\v cenko-Pastur density $\rho_N^\beta$ (defined in Eq. \eqref{dsd_beta}) for large but finite 
$N,M$ with $N/M=q\in (0;1]$.

We are therefore interested in computing functionals of the form $\langle \int f(\lambda) \rho_N^\beta( d\lambda)\rangle$ where $f$ is a test function and where 
 $(\lambda_1,\lambda_2,\dots,\lambda_N)$ is distributed according to the jpdf $P_\beta$ defined in Eq. \eqref{P_beta} for $\beta>0$. The idea is to use the
stochastic process $\lambda_1(t)\leq \dots \leq \lambda_N(t)$ following the SDE \eqref{SDE_ev} that converges in law when $t\rightarrow \infty$ to the vector
$(\lambda_1\leq \dots \leq \lambda_N)$ distributed according to $P_\beta$.  

Note that in order to have a well behaved limiting spectral distribution with edges that do not depend on $M$ or on $\beta$, we will choose in this section $\sigma=1/\sqrt{M\beta}$
(or alternatively rescale all eigenvalues by $1/M \beta$).

Using again It\^o's formula for $\int f(\lambda) \rho_N^t( d\lambda)$ (where $\rho_N^t$ is still defined by Eq. \eqref{emp_meas_lambda_t}) and Eq. \eqref{SDE_ev}, we obtain 
\begin{align}
\label{ito_f}
{ d } \int f(x) \rho_N^t(dx) &=  \int \left(- x + 1 \right) f'(x) \rho_N^t(dx)  dt
+ \frac{1}{M\beta} (2-\beta) \int x f''(x) \rho_N^s(dx) dt \\
 &+ \frac{ q}{2}  \int \int \frac{f'(x)-f'(y)}{x-y} (x+y) \rho_N^t(dx) \rho_N^t(dy) dt + { d}M_t^N   \notag
\end{align}
where $dM_t^N = \frac{1}{N} \sum_{i=1}^N 2 \sqrt{\frac{\lambda_i}{M\beta}}\, f'(\lambda_i) db_i$ is a noise term of variance 
$\frac{4}{q \beta M^2} \int \lambda f'(\lambda)^2\rho_N^t({ d}\lambda) { d}t$.

Let us introduce the Stieltjes transform $H_t(z)$ of the probability measure $\rho_N^t$ defined as 
\begin{equation*}
H_t(z) = \int \frac{\rho_N^t({ d}x)}{x-z} \,.
\end{equation*} 
We now apply \eqref{ito_f} to the particular function $f(x) =1/(x-z)$ and we take the expectation with respect to the $b_i$; Eq. \eqref{ito_f} rewrites as 
\begin{align}
\label{eq_H_t}
\frac{\partial \langle H_t \rangle }{\partial t} &= \left[\langle H_t\rangle + z \frac{\partial \langle H_t \rangle}{\partial z}\right] -  
 \frac{\partial \langle H_t \rangle}{\partial z} +  \frac{1}{M\beta} (2-\beta) \left[2  \frac{\partial \langle H_t \rangle}{\partial z}+ z  
\frac{\partial^2 \langle H_t\rangle }{\partial z^2} \right] \\
&+   q \left[ \frac{\partial \langle H_t \rangle }{\partial z} + \langle H_t \rangle^2 + 2 z \langle H_t \rangle  \frac{\partial \langle H_t \rangle}{\partial z} \right]
+  q \left[  \langle H_t^2 \rangle -   \langle H_t \rangle^2 \right] 
+ z  q   \frac{\partial }{\partial z} \left[ \langle H_t^2 \rangle -  \langle H_t \rangle^2 \right] \notag
\end{align} 
where $\langle \cdot\rangle$ denotes the expectation with respect to the Brownian motions $b_i$. The two last terms come from the replacement of $\langle H_t^2\rangle$
by $\langle H_t\rangle^2$ in the third term of the right hand side of \eqref{eq_H_t}. By setting 
\begin{equation}
F_t(z) = \langle H_t(z)^2\rangle - \langle H_t(z)\rangle^2\,,
\end{equation}
Eq. \eqref{eq_H_t} can be rewritten as 
\begin{align}\label{eq_H_t.1}
\frac{\partial \langle H_t \rangle }{\partial t} &= \left[\langle H_t\rangle + z \frac{\partial \langle H_t \rangle}{\partial z}\right] -  
\frac{\partial \langle H_t \rangle}{\partial z} +  \frac{1}{M\beta} (2-\beta) \left[2  \frac{\partial \langle H_t \rangle}{\partial z}+ z  
\frac{\partial^2 \langle H_t\rangle }{\partial z^2} \right] \\
&+   q \left[ \frac{\partial \langle H_t \rangle }{\partial z} + \langle H_t \rangle^2 + 2 z \langle H_t \rangle  \frac{\partial \langle H_t \rangle}{\partial z} \right]
+  q \left[F_t + z  \frac{\partial F_t }{\partial z}  \right] \notag\,.
\end{align} 
To simplify notations, we will now omit the $\langle\cdot\rangle$ and write $H$ instead of $\langle H\rangle$.
The stationary solution of \eqref{eq_H_t.1} writes simply as 
\begin{align}
 \left[H+ z \frac{{ d}H}{{ d}z} \right] -    \frac{{ d}H}{{ d}z} +  \frac{1}{M\beta} (2-\beta) \left[2  \frac{{ d}H}{{ d}z}+ z  \frac{{ d}^2H}{{ d}z^2}  \right] \\ 
+   q \left[ \frac{{ d}H}{{ d}z}+ H^2 
+ 2 z H \frac{{ d}H}{{ d}z} \right] + q \left[F+ z  \frac{{ d} F}{{ d} z}  \right] =0 \notag\,.
\end{align} 
which can be rewritten as 
\begin{align}\label{eq_H_1}
q H(z) [H(z) + 2 z  \frac{{ d}H}{{ d}z}] + \frac{1}{2} [H(z)+2z\frac{{ d}H}{{ d}z}] + \frac{1}{2} H(z) 
+ \frac{1}{2M\beta}(2-\beta) [3 \frac{{ d}H}{{ d}z} + 2z \frac{{ d}^2H}{{ d}z^2}] \\ 
+   \left[\frac{1}{2M\beta}(2-\beta)+  (-1+q)\right] \frac{{ d}H}{{ d}z}
+  q \left[F+ z  \frac{{ d} F}{{ d} z}  \right]  = 0\,.\notag
\end{align}

Eq. \eqref{eq_H_1} can be integrated easily by doing the change of function $G(z) = z H(z^2)$. 
Indeed, we just need to write \eqref{eq_H_1} for $z^2$ instead of $z$ and then multiply the corresponding equation by $z$ to obtain the following equation
\begin{align}\label{eq_G_1}
q G &\frac{{ d}G}{{ d}z} +  \frac{1}{2}\left[z \frac{{ d}G}{{ d}z} + G\right] +  \frac{1}{4M\beta}(2-\beta)  \frac{{ d}^2G}{{ d}z^2} \\ 
&+  \frac{1}{2}\left[\frac{1}{2M\beta}(2-\beta)+  (-1+q)\right]  2z \frac{{ d}H}{{ d}z}(z^2) +  q z \left[F(z^2)+ z^2  \frac{{ d} F}{{ d} z}(z^2)  \right]  = 0\,.\notag
\end{align}
Equation \eqref{eq_G_1} can be straightforwardly integrated with respect to $z$ as 
\begin{equation}\label{eq_G_2}
q G^2 +   z G + \frac{1}{2M\beta}(2-\beta)  \frac{{ d}G}{{ d}z} + \left[\frac{1}{2M\beta}(2-\beta)+ \beta (-1+q)\right] \frac{G(z)}{z} +  q z^2 F(z^2)= -1
\end{equation}
where the integration constant is chosen so that $z G(z) \sim -1$ when $|z|\rightarrow \infty$. 
Note that the asymptotic behavior for $H$ is therefore also $zH\sim -1$ when $|z|\rightarrow \infty$ as is expected for the Stieltjes transform of a probability measure. 
Rewriting now Equation \eqref{eq_G_2} in term of the function $H$, we obtain 
\begin{equation}\label{eq_dif_H}
q H^2 + H \left[ 1+ \frac{1}{z} \left(q-1+\frac{1}{M} \frac{2-\beta}{\beta}\right) \right] + \frac{1}{M} \frac{2-\beta}{\beta} \frac{{ d}H}{{ d}z} + \frac{1}{z}
+ q \, F(z) = 0\,.
\end{equation}

Now, using the result about the noise in the Mar\v cenko-Pastur law obtained in 
\cite{lytova} by Lytova and Pastur, we know that in the limit of large $M$, we have
for $\beta=1$ or $2$, 
\begin{align}\label{res_lytpast}
&F(z) \sim_{M\rightarrow \infty} \\ 
& \frac{1}{M^2} \frac{1}{q^2} \frac{1}{2\beta\pi^2} \int_{\gamma_-}^{\gamma_+} \int_{\gamma_-}^{\gamma_+} \frac{d\lambda d\mu}{(\lambda-z)^2(\mu-z)^2} 
 \frac{4q - (\lambda-(1+q))(\mu-(1+q))}{\sqrt{4q - (\lambda-(1+q))^2}\sqrt{4q- (\mu-(1+q))^2}}  \notag
 \end{align}
with $\gamma_\pm = (1 \pm \sqrt{q})^2$.

The idea to obtain the correction to the Mar\v cenko-Pastur law is to use perturbation theory in Eq. \eqref{eq_dif_H}.
More precisely, we want to compute explicitly the coefficients $\rho_0,\rho_1$ and $\rho_2$ such that the eigenvalue density
of a $\beta$-Wishart matrix writes under the form, in the limit of large $N,M$ with $N/M=q$, 
\begin{equation}\label{expans_rho}
\rho(\lambda) = \rho_0(\lambda) + \frac{1}{M} \rho_1(\lambda) + \frac{1}{M^2} \rho_2(\lambda) + o\left(\frac{1}{M^2}\right)\,.
\end{equation}

Note that this asymptotic expansion \eqref{expans_rho} is obtained by perturbation theory and therefore is valid only for the values of $\lambda$ such that 
the correction terms $\rho_1(\lambda)/M$ and $\rho_2(\lambda)/M^2$ are negligible compared to the leading term $\rho_0(\lambda)$ in the limit of large $M$, i.e. 
for the values of $\lambda$ such that $\rho_0(\lambda)\neq 0$. 
The expansion \eqref{expans_rho} is not valid outside the Mar\v cenko-Pastur sea, i.e.,
it breaks down near the edges (see below). In addition, here we are talking about
smoothed density, hence it contains no oscillatory term in the finite $N$ bulk 
corrections~\cite{FFG}. 

To this purpose, we first write $H(z)$ under the form  
\begin{equation}\label{pth_H}
H(z) = H_0(z) + \frac{1}{M} H_1(z) + \frac{1}{M^2} H_2(z) + o\left(\frac{1}{M^2}\right)
\end{equation}
and we plug Eq. \eqref{pth_H} into Eq. \eqref{eq_dif_H}. By solving the equation to leading order, we find the following expression for $H_0(z)$ 
\begin{align}\label{expr_H_0}
H_0(z) &= \frac{1}{2q} \frac{-(z+q-1)+ \sqrt{(z-\gamma_-)(z-\gamma_+)}}{z} \\
&=  \frac{1}{2q} \frac{-(z+q-1)+ \sqrt{(z-(1+q))^2 - 4 q }}{z} \,.\notag
\end{align} 
We deduce from this the famous Mar\v cenko-Pastur result: the eigenvalue density converges in the limit of large $N,M$ with $N/M=q$ to the Mar\v cenko-Pastur density 
as expected given by 
\begin{align*}
\rho_0(\lambda) = \frac{1}{2\pi q} \frac{ \sqrt{(\lambda-\gamma_-)(\gamma_+-\lambda)}}{\lambda}\,. 
\end{align*}
For all value of $\beta >0$, we can now compute the $1/M$ correction to the Mar\v cenko-Pastur density
by plugging Eq. \eqref{pth_H} into Eq. \eqref{eq_dif_H} and solve to order $1/M$. This gives the following expression for $H_1(z)$ 
\begin{align*}
H_1(z) = - \left(\frac{1}{\beta} - \frac{1}{2} \right) \frac{1}{q}\left[\frac{1}{2} \left(\frac{1}{z-\gamma_+} + \frac{1}{z-\gamma_-}\right) -\frac{1}{\sqrt{(z-\gamma_+)(z-\gamma_-)}}  \right]
\end{align*}
and the corresponding $1/N$ correction to the density is then given (for all $\beta$) by 
\begin{equation}
\rho_1(\lambda) = \left(\frac{1}{\beta} - \frac{1}{2} \right) \frac{1}{q} \left[\frac{1}{2} \left( \delta(\lambda-\gamma_+) +\delta(\lambda-\gamma_-)\right) 
- \frac{1}{\pi} \frac{{ d}\lambda}{\sqrt{(\lambda-\gamma_-)(\gamma_+-\lambda)}} \right]\,.
\end{equation}
Comparing $H_1(z)/M$ with $H_0(z)$, we see that the correction term ceases to be negligible when $|\lambda-\gamma_{\pm}| \sim M^{-2/3}$, as expected: this is indeed the standard edge 
scaling that defines the Tracy-Widom region. Note that this $\mathcal{O}(1/N)$ correction term
was derived earlier [\cite{FFG} and references therein] for $\beta=1$, $2$ and $4$, but
our result is valid for general $\beta$.

For the particular value $\beta=1$ or $2$, we can use the result of Lytova and Pastur 
stated above in Eq. \eqref{res_lytpast} to compute the $1/M^2$ correction with the same method
by solving the equation until order $1/M^2$.

Let us first compute an explicit expression for $F(z)$ from the integral representation in Eq. \eqref{res_lytpast}  
\begin{align*}
 q^2 M^2 F(z) &= \frac{1}{2\beta\pi^2} \Bigg[  4q \left( \int_{\gamma_-}^{\gamma_+}  \frac{1}{(\lambda-z)^2} \frac{{ d}\lambda}{\sqrt{4q - (\lambda-(1+q))^2}} \right)^2
\\&  -  \left( \int_{\gamma_-}^{\gamma_+} \frac{{ d}\lambda}{(\lambda-z)^2}   \frac{\lambda-(1+q)}{\sqrt{4q - (\lambda-(1+q))^2}} \right)^2\Bigg] + o(1) \\
&=  \frac{1}{2\beta\pi^2} \Bigg[  4q \left( - \frac{\pi}{(z-\gamma_-)(z-\gamma_+)} \frac{2z-\gamma_- - \gamma_+}{\sqrt{(z-\gamma_-)(z-\gamma_+)}}  \right)^2 \\
&- \frac{\pi^2 }{(z-\gamma_-)(z-\gamma_+)} \left(1 - \frac{1}{2} \frac{(2z-\gamma_- -\gamma_+)^2}{(z-\gamma_-)(z-\gamma_+)} \right)^2 \Bigg] +o(1)\\
&=\frac{1}{2\beta} \frac{1}{(z-\gamma_-)(z-\gamma_+)} \left[ 4q  \frac{(2z-\gamma_- - \gamma_+)^2}{(z-\gamma_-)^2(z-\gamma_+)^2} -
 \left(1 - \frac{1}{2} \frac{(2z-\gamma_- -\gamma_+)^2}{(z-\gamma_-)(z-\gamma_+)} \right)^2 \right]+o(1)\\
 &= \frac{1}{2\beta} \frac{1}{(z-\gamma_-)(\gamma_+ - z)} + o(1)\,.
\end{align*}

Then we can turn to compute $H_2(z)$ and deduce from this computation the expression for $\rho_2(\lambda)$
\begin{align*}
\rho_2(\lambda) &= - 2\left(\frac{1}{2} - \frac{1}{\beta} \right)^2 \frac{1}{q}  \frac{1}{\sqrt{(\lambda-\gamma_-)(\gamma_+-\lambda)}} 
\Bigg[ \frac{1}{2}\left( \frac{1}{\lambda-\gamma_-} + \frac{1}{\lambda-\gamma_+}  \right) \\ 
& - \frac{\lambda}{2} \left(\frac{1}{(\lambda-\gamma_-)^2} + \frac{1}{(\lambda-\gamma_+)^2} \right) \Bigg] 
+  \frac{1}{2q\beta} \frac{\lambda}{(\lambda-\gamma_-)^{3/2}(\gamma_+ - \lambda)^{3/2}}  \,.
\end{align*}

Again, the comparison of this correction term with the dominant term indicates that our 
perturbation expansion breaks down when $|\lambda-\gamma_{\pm}| \sim M^{-2/3}$.


\section{Conclusions}\label{concl}

In summary, we proposed a random matrix model (invariant under similarity transformations) whose joint density of eigenvalues is given by the classical $\beta$-Wishart ensemble
where the quantization of the Dyson index $\beta$ is lifted. The procedure is constructive and is described in section \ref{diffmat}. The resulting ensemble is by construction
invariant under similarity transformations with Haar distributed eigenvectors 
The diffusive evolution equation
for the eigenvalues involves the Dyson index of the ensemble as a free parameter. Letting it scale with the size $M$ of the matrix, the spectral density of the ensemble
becomes a one-parameter continuous family interpolating between the familiar Mar\v cenko-Pastur distribution and a certain type of Gamma distribution. On the other hand, keeping the
Dyson index unscaled but not quantized, we showed that a careful analysis of the full Stieltjes transform equation lead naturally to $1/N$ and $1/N^2$ corrections (and possibly systematically to any order)
to the average spectral density (Mar\v cenko-Pastur) for all $\beta\neq 2$. This then
extends the previous work~\cite{FFG,Forrester12} on the $\mathcal{O}(1/N)$ correction term for 
$\beta=1$, $2$ and $4$.
To order $O(1/N^2)$, our result is valid for $\beta=1$ and $\beta=2$. It would be
interesting to see (or conjecture) if this formula to $O(1/N^2)$ term is valid for general
$\beta$. 

In this work, we have computed the crossover density as a function of the interpolating 
parameter $c$. In the limit $c\to \infty$, it reduces to the standard Mar\v cenko-Pastur 
density, whereas the opposite limit $c\to 0$ corresponds to the Gamma laws associated 
with the CIR process. It would be interesting to extend our analysis to the distribution
of the largest eigenvalue. As in the case of bulk density, we would expect
a $c$-dependent distribution for the largest eigenvalue, properly centered and scaled, 
interpolating between
the Tracy-Widom distribution ($c\to \infty$ limit) and Gumbel 
distribution (as $c\to 0$).  

%

\appendix 

\section{Derivation of \eqref{P_star}}\label{deriv_FP}
The Fokker-Planck equation for the transition probability density $P(\lambda_1,\cdots,\lambda_N;t)$ of the 
process $(\lambda_1(t),\cdots,\lambda_N(t))$ which satisfies the stochastic differential system \eqref{sde_ev_p} reads 
\begin{equation}\label{eq_FP}
\frac{\partial P}{\partial t} = - \sum_{i=1}^{N} \frac{\partial }{\partial \lambda_i}\left[ P \left(-\lambda_i+ pM + (1-\delta) p + 
p \sum_{k\neq i} \frac{\lambda_i+\lambda_k}{\lambda_i-\lambda_k} \right)
 \right] +  2 \sum_{i=1}^N \frac{\partial^2}{\partial \lambda_i^2} \left[ \lambda_i P \right] \,.
\end{equation}
The stationary solution is the solution which does not depend on time $t$, satisfying
\begin{equation}
 - \sum_{i=1}^{N} \frac{\partial }{\partial \lambda_i}\left[ P \left(-\lambda_i+ pM + (1-\delta) p + 
p \sum_{k\neq i} \frac{\lambda_i+\lambda_k}{\lambda_i-\lambda_k} \right)
 \right] +  2 \sum_{i=1}^N \frac{\partial^2}{\partial \lambda_i^2} \left[ \lambda_i P \right]  =0 \,.
\end{equation}
It is easy to check using elementary algebra that the jpdf $P^*$ defined in \eqref{P_star} verifies Eq. \eqref{eq_FP} as in fact we can verify that for all $i$, 
\begin{equation}
2  \frac{\partial}{\partial \lambda_i} \left[ \lambda_i P^* \right]  =  P^* \left(-\lambda_i+ pM + (1-\delta) p + 
p \sum_{k\neq i} \frac{\lambda_i+\lambda_k}{\lambda_i-\lambda_k} \right)  \,. 
\end{equation} 
 
\section{Algorithmic description of how to build the process ${\bf W}_t^n$ in practice}\label{algo}

Let us describe shortly an algorithmic description of how to build the process ${\bf W}_t^n$ in practice, on a discrete grid. 
First note that this algorithmic description needs a discrete grid and that it does not reproduce exactly the process ${\bf W}_t^n$ 
but only a discretized approximation of it. 
Choose a large value of $n$ and 
an initial symmetric matrix ${\bf W}_0$. 
The construction is iterative. Suppose that the process is constructed until time $k/n$ and let us explain how to compute the matrix ${\bf W}_{(k+1)/n}^n$ at 
the next discrete time of the grid, $(k+1)/n$. 
\begin{enumerate}
\item Step 1.  We first need to compute the matrix $\sqrt{{\bf W}_{k/n}^n}$. It suffices to compute the orthogonal matrix ${\bf O}_{k/n}^n$ such that
\begin{equation*}
{\bf W}_{k/n}^n = {\bf O}_{k/n}^n {\bf\Sigma}_{k/n}^n {{\bf O}_{k/n}^n}^\dagger
\end{equation*} 
where ${\bf\Sigma}_{k/n}^n$ is the diagonal matrix composed of the eigenvalues of ${\bf W}_{k/n}^n$ (in increasing order). 
The eigenvalues of the matrix ${\bf W}_{k/n}^n$ should be non negative as the eigenvalues process of ${\bf W}_t^n$ are almost surely non negative at all time $t$. 
However, due to the discretization scheme necessary for algorithmic procedure, the non-negativity can fail.
To avoid this problem, we define $\sqrt{{\bf W}_{k/n}^n}$ as
\begin{equation}
\sqrt{{\bf W}_{k/n}^n} = {\bf O}_{k/n}^n \sqrt{{\bf\Sigma}_{k/n}^n} {{\bf O}_{k/n}^n}^\dagger
\end{equation}
where $\sqrt{{\bf\Sigma}_{k/n}^n}$ is the diagonal matrix composed of the square roots of the absolute values of the eigenvalues of ${\bf W}_{k/n}^n$ 
 (again in increasing order).
\item Step 2. We sample the Bernoulli random variable $\epsilon_k^n$ with $\P[\epsilon_k^n=1] = p = 1-\P[\epsilon_k^n=0]$. 
\item Step 3. It depends on the value of $\epsilon_k^n$:
\begin{itemize}
\item if $\epsilon_k^n=1$, we sample a $N \times N$ matrix ${\bf G}_n$ filled with independent Gaussian variables with mean $0$ and variance $1/n$ and then 
we compute the matrix ${\bf W}_{(k+1)/n}^n$ by the formula 
\begin{equation*}
{\bf W}_{(k+1)/n}^n = \left(1-\frac{1}{n}\right) {\bf W}_{k/n}^n + \sqrt{{\bf W}_{k/n}^n} \, {\bf G}_n+ {\bf G}_n^\dagger \, \sqrt{{\bf W}_{k/n}^n} + \frac{1}{n} M\,  \,{\bf I} \,.
\end{equation*}
\item if $\epsilon_k^n=0$, we sample $N$ independent Gaussian variables $(z_1,\cdots,z_N)$ with mean $0$ and variance $1/n$. We then compute 
the matrix ${\bf Y}_n$, which is co diagonalizable 
with the matrix ${\bf W}_{k/n}^n$, defined as the product
\begin{equation}
{\bf Y}_n:= {\bf O}_{k/n}^n  {\rm Diag}\left(z_1,z_2,\dots,z_N\right) {{\bf O}_{k/n}^n}^\dagger\,.
\end{equation} 
Finally we obtain the matrix ${\bf W}_{(k+1)/n}^n$ by
\begin{equation*}
{\bf W}_{(k+1)/n}^n = \left(1-\frac{1}{n}\right) {\bf W}_{k/n}^n + \sqrt{{\bf W}_{k/n}^n} \, {\bf Y}_n+ {\bf Y}_n^\dagger \, \sqrt{{\bf W}_{k/n}^n} + \frac{1}{n} \delta \,  \,{\bf I} \,.
\end{equation*}
\end{itemize}
\end{enumerate}

{\bf Acknowledgments}

We would like to thank A. Comtet and A. Guionnet for useful discussions. SNM acknowledges support by ANR grant 2011-BS04-013-01 WALKMAT.

\end{document}